# Big Data Computing Using Cloud-Based Technologies: Challenges and Future Perspectives

Samiya Khan[1], Kashish A. Shakil[2] and Mansaf Alam[3]

[1]samiyashaukat@yahoo.com, [2]shakilkashish@yahoo.co.in, [3]malam2@jmi.ac.in

Department of Computer Science, Jamia Millia Islamia, New Delhi

***Abstract***: *The excessive amounts of data generated by devices and Internet-based sources at a regular basis constitute, big data. This data can be processed and analyzed to develop useful applications for specific domains. Several mathematical and data analytics techniques have found use in this sphere. This has given rise to the development of computing models and tools for big data computing. However, the storage and processing requirements are overwhelming for traditional systems and technologies. Therefore, there is a need for infrastructures that can adjust the storage and processing capability in accordance with the changing data dimensions. Cloud Computing serves as a potential solution to this problem. However, big data computing in the cloud has its own set of challenges and research issues. This chapter surveys the big data concept, discusses the mathematical and data analytics techniques that can be used for big data and gives taxonomy of the existing tools, frameworks and platforms available for different big data computing models. Besides this, it also evaluates the viability of cloud-based big data computing, examines existing challenges and opportunities, and provides future research directions in this field.*

---

**1.0 Introduction**

The advent of Internet and rapidly increasing popularity of mobile and sensor technologies have led to an outburst of data in the systems and web world. This data explosion has posed several challenges to systems, traditionally used for data storage and processing. In fact, the challenges are so grave that it would not be wrong to state that traditional systems can no longer fulfill the growing needs of data-intensive computing.

The two main requirements of big data analytics solutions are (1) scalable storage that can accommodate the growing data (2) high processing ability that can run complex analytical tasks in finite and allowable time. Among many others, the Cloud Computing technology is considered an apt solution to the requirements of big data analytics solutions considering the scalable, flexible and elastic resources that it offers (Philip Chen and Zhang 2014). Firstly, the Cloud offers commodity machines that provide scalable yet cost-effective storage solutions. Besides this, the processing ability of the system can be improved by adding more systems dynamically to the cluster. Therefore, the flexibility and elasticity of the Cloud are favorable characteristics for big data computing.

Cloud-based big data analytics technology finds a place in future networks owing to the innumerable 'traditionally unmanageable abilities and services' that this technology offers. The general definition of future networks describe it as a network that possesses the capabilities to provide services and facilities that existing network technologies are unable to deliver. Therefore, a component network, an enhanced version of an existing network or a federation of new and existing networks that fulfill the above-mentioned requirements can be referred to as future networks.

The applicability of the big data concept and its relevance to society and businesses alike makes it a potential game changer in the technological world so much so that many people consider this concept just as important to businesses and society as the Internet. This brings us to look at reasons why big data needs to be studied and researched. The answer to this question lies in fundamental principles of statistical science. One of the prerequisites of any statistical analysis is data. Moreover, the higher the number of samples, the better is the computed analysis for the given data. Therefore, more data directly implies better analyses, which in turn means better decision-making.

From an organizational perspective, efficient decision-making can have a significant impact on improving the operational efficiency and productivity of the organization. This notion can be scaled down to the individual level and the availability of better analyses in the form of applications and systems can increase individual productivity and efficiency, manifold. Evidently, the big data concept is capable of bringing about a revolution in the society and business world and can change the way we live our lives, just like the Internet did, years ago.

This technology finds applications in diverse fields and areas. Although, the big data problem can model any data-intensive system, there are some established practical applications that have gained popularity amongst the research community and governing authorities. These applications include smart cities (Khan, Anjum, and Kiani 2013), analytics for healthcare sector (Raghupathi and Raghupathi 2014), asset management system for railways (Thaduri, Galar, and Kumar 2015), social media analytics (Burnap et al. 2014), geospatial data analytics (Lu et al. 2011), customer analytics for the banking sector (Sun et al. 2014), e-commerce recommender systems (Hammond and Varde 2013) and Intelligent Systems for transport (Chandio, Tziritas, and Xu 2015), in addition to several others.

This chapter shall illustrate the big data problem, giving useful insights in the tools,

techniques and technologies that are currently being used in this domain, with specific reference to Cloud Computing, as the infrastructural solution for the storage and processing requirements of big data. Section 1 provides a comprehensive definition of big data and the two most popular models used for big data characterization namely Multi-V model (Section 1.1) and HACE Theorem (Section 1.2). The lifecycle of big data and the different processes involved have been elaborated upon in Section 2. In order to process and analyze big data, several existing mathematical techniques and technologies can be used. Section 3 discusses the techniques used for big data processing. This section has been divided into two sub-sections namely mathematical techniques (Section 3.1) and data analytics techniques (Section 3.2). The techniques under each of these sub-headings have been described.

Big data has found applications in diverse fields and domains. In order to fulfill the varied requirements of big data applications, six computing models exist. Each of these computing models serve different computing requirements of distinctive big data applications, which include batch processing (Section 4.1), stream processing (Section 4.2), graph processing (Section 4.3), DAG processing (Section 4.4), interactive processing (Section 4.5) and visual processing (Section 4.6). The tools available for each of these computing models have been described in the sub-sections.

The techniques and computing models need to be efficiently implemented for big data, which require the systems to provide magnified storage and computing abilities. The adaptation of standard techniques for the big data context requires technologies like Cloud Computing, which have been described in Section 5. This section shall examine how big data analytics can use the characteristics of Cloud Computing for optimal benefit to several real world applications. The first few sub-sections introduce Cloud Computing and describe characteristics (Section 5.1), delivery models (Section (5.2), deployment models (Section 5.3) and how Cloud Computing is the most appropriate technology for big data (Section 5.4). In view of the fact that Hadoop is the most popular computing framework for big data, Hadoop on the Cloud (Section 5.5) gives the implementation options available for moving Hadoop to the Cloud. Section 6 summarizes the challenges identified for big data computing in the cloud environment and Section 7 discusses directions for future research.

## 2.0 Defining Big Data

Several definitions for big data exist owing to the varied perspectives and perceptions with which this concept is viewed and understood. Regardless of the source of a definition, most big data experts possess a unanimous viewpoint on the fact that big data cannot be restricted only to the dimension of volume. Many more dimensions, some of which may be application or data-source dependent, need to be explored before a comprehensive definition for big data can be articulated.

The most accepted definition of big data describes it as huge volumes of exponentially growing, heterogeneous data. Doug Laney of Gartner, in the form of the 3V Model, gave the first definition of big data (Gartner 2016). The fundamental big data characteristics included in this classification are volume, variety and velocity. However, big data, as a technology, picked up recently after the availability of open source technologies like NoSQL (Salminen 2012) and Hadoop[1], which have proven to be effective and efficient solutions for big data storage and processing. Accordingly, the definition of big data was modified to data that cannot be stored, managed and processed by traditional systems and technologies.

---

[1] http://hadoop.apache.org/

To enhance the technical precision of existing definitions, Shaun Connolly introduced the terms transactions, interactions and observations, which were added to the big data definition (Samuel 1959). While 'transactions' is a term used to describe data that has already been collected and analyzed in the past, 'interactions' include data that is collected from things and people. A class of data missed by both these categories is the data that is automatically collected and constitutes 'observations'. Barry Devlin gave a similar, yet clearer definition of big data, describing it in terms of machine-generated data, human-sourced data and process-mediated data (Ratner 2003).

From the business-effectiveness perspective, transaction data hold little relevance in view of the fact that it is old data and by the time, it is collected and analyzed, the results become obsolete or lesser relevant. On the contrary, new data needs to be analyzed efficiently to provide predictions, which can be used to make timely interventions. Sentiment analysis is an application that works on this perspective and uses big data as signals. This was formalized into a timing and intent-based classification of big data (Buyya 2016).

Besides the above mentioned, there have been several other viewpoints and perspectives on big data. Scholars like Matt Aslett sees big data as an opportunity to explore the potential of data that was previously ignored due to limited capabilities of traditional systems while some others call it a new term for old applications and technologies like Business Intelligence (Abu-Mostafa, Magdon-Ismail, and Lin 2012). Regardless of the big data definition one chooses to follow, nothing can take away the fact that big data opens doors to unlimited opportunities and in order to make use of this reserve, we need to develop new or modify the existing tools and technologies.

### *2.1 Multi-V Model*

The initial 3-V model, given by Doug Laney in the year 2001, includes volume, variety and velocity, as the three fundamental big data characteristics (Gartner 2016). The amount of data included in a dataset indicates the volume of data, which is the most obvious characteristic of big data. The data concerned may come from different sources and can be of diverse types. For instance, data coming from a social media portal includes textual data, audio and video clips and metadata. In order to accommodate for these different types of data, big data is said to include structured data, semi-structured and unstructured data. Lastly, data may be produced in batches, near time or real-time. The speed at which the concerned data is being generated denotes the velocity characteristic of big data. Figure 1 depicts the scope of interest for the three V's mentioned above.

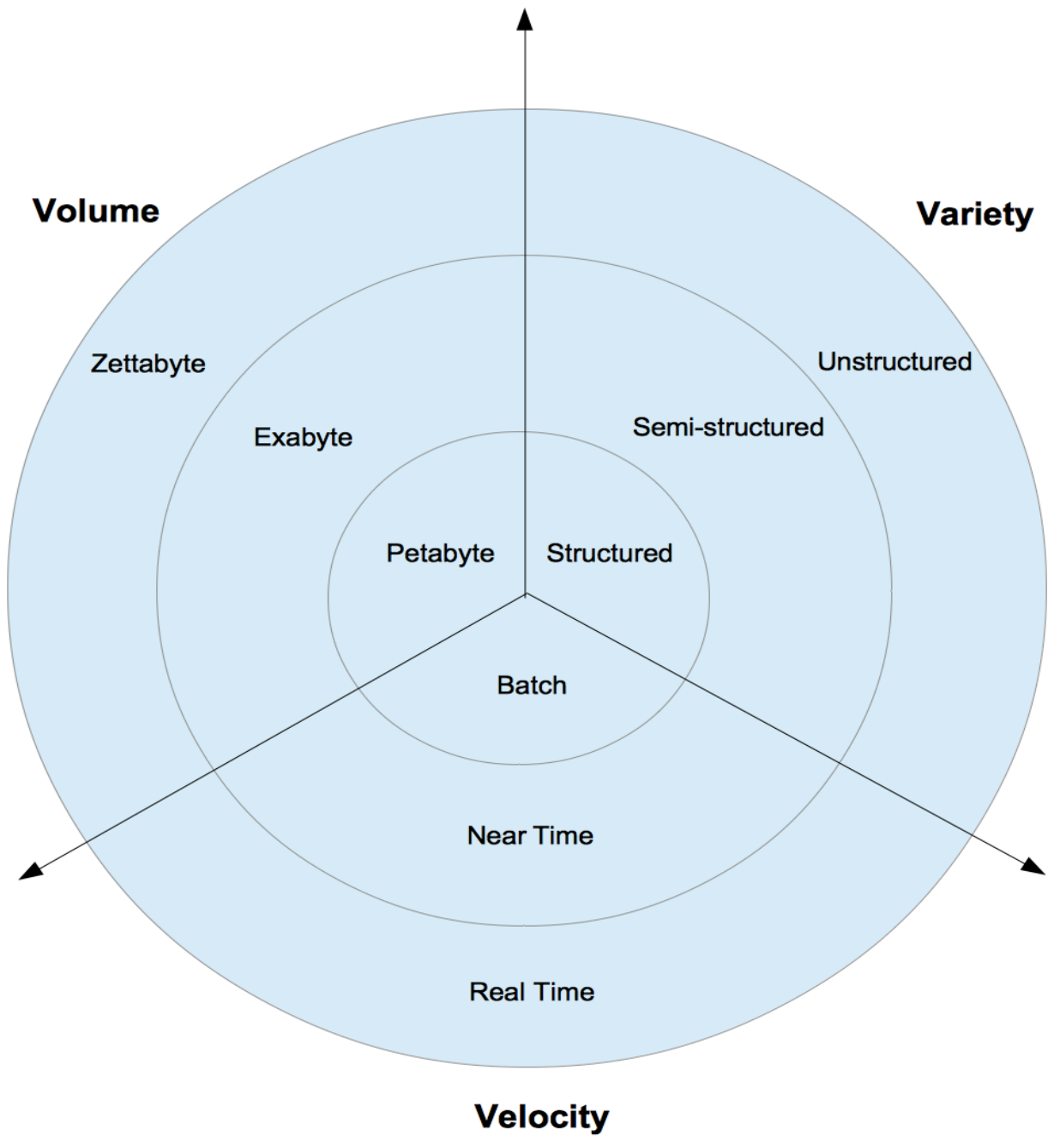

Fig. 1 – Volume, Variety and Velocity of Big Data

However, this initial model has been expanded in different dimensions by including V's like veracity, variability and value, in addition to several others. One of the most recent adaptations of this model is the $3^2V$ model (Buyya 2016). This model divides the V's into three classes namely, data (volume, variety and velocity), business intelligence (value, visibility and verdict) and statistics (veracity, variability and verdict). Each of these classes include 3V's specific to their domain and have been described in Table 1.

Table 1 – Description of $3^2V$ Characteristics

| Class | V Characteristic | Description |
|---|---|---|
| Data Domain | Volume | Amount of data |
| | Variety | Types of data included |
| | Velocity | Speed of data generation |
| Business Intelligence Domain | Value | The business value of the information present in the data |
| | Visibility | Insight, hindsight and foresight of the problem and its adequate solutions |
| | Verdict | Possible decision to be made on the basis of the problem's scope, computational capacity and available resources |
| Statistics Domain | Veracity | Uncertainty and trustworthiness of data |
| | Variability | Variations and complexity of data |
| | Validity | Objectivity with which the data is collected |

## *2.2 HACE Theorem*

HACE stands for Heterogeneous, Autonomous, Complex and Evolving (Xindong et al. 2014) and describes big data as a large volume of heterogeneous data that comes from autonomous sources. These sources are distributed in nature and the control is essentially decentralized. This data can be used for exploration of complex and evolving relationships. These characteristics make identification and extraction of useful information from this data, excessively challenging. In order to ensure a clear understanding, a detailed description of these characteristics has been given below.

1. Huge Volumes of Data from Heterogeneous Sources

   Volume, as identified by many different modeling frameworks, is one of the fundamental characteristics of big data. In addition to this, with the increasing popularity of Internet and social networking portals, data is no longer confined to a format or source. Diverse sources of data like Twitter, Facebook, LinkedIn, organizational repositories and other external and internal data sources result in data multi-dimensionality. Data representation may also vary as a result of varying methods of data collection, data collector preferences and application-specific needs of the system.

2. Distributed, Autonomous Sources with Decentralized Control

   The nature of the sources is usually autonomous and distributed. As a result, the production and collection of data does not require connection to any central control. Many parallels can be drawn between the World Wide Web and big data technology at this level. Just like the web servers that possess information of their own and can function independently, the sources of big data are also independent entities that do not require any external monitoring or control. More specifically, data sources are guided by local government regulations and market parameters, which result in restructuring of data representations. The distributed nature of sources increases fault tolerance, which is one of the significant advantages of such a system.

3. Complex and Evolving Nature of Relationships

   Any form of data representation requires exploration of associations and relationships between the different entities involved. Keeping the dynamic nature of this world in view, several spatial, temporal and other kinds of factors are involved in representing entities and their continuously evolving relationships. Owing to the presence of different data types like audio, video, documents and time-series data, in addition to several others, this data possesses high complexity.

**3.0 Big Data Lifecycle**

The lifecycle of big data includes several phases, which include data generation, acquisition, storage and processing of data. These four phases have been explained below and illustrated in Fig. 2.

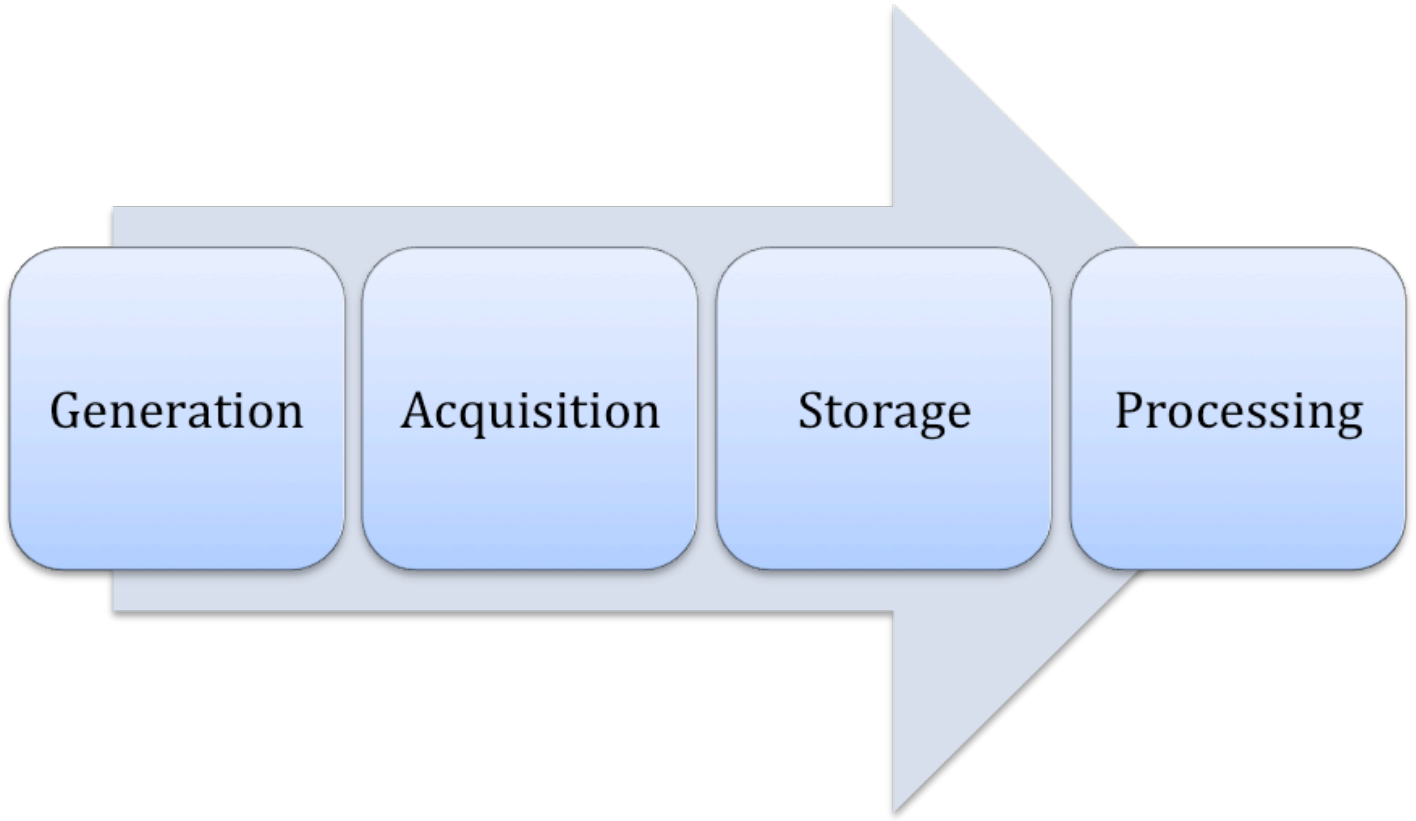

Fig. 2 – Big Data Lifecycle

1 Generation

Data is the most rapidly increasing resource in the world. Perhaps, the reason for this staggering rise in its generation is the diverse types of devices, entities and systems involved. With the rapid advancement in technology, devices like sensors, online portals, social networking websites and online systems like online trading and banking, in addition to many others, have come into existence. All these systems, portals and devices generate data at a periodic basis, contributing to the volume, variety and velocity of big data.

2 Acquisition

Now that we know that big data is being generated by diverse sources, this data needs to be acquired by big data systems for analysis. Therefore, during this stage of the big data lifecycle, the raw data generated in the world is collected and given to the next stage for further processing. Examples of data acquisition systems include log files, sensing systems, web crawlers and REST APIs provided by portals. Since big data includes different types of data, an efficient pre-processing mechanism is required. Common methods used for this purpose include data cleaning, redundancy reduction and data integration. This is a crucial step for ensuring data veracity.

3 Storage

The sheer volume of big data overwhelms traditional storage solutions. In order to address the challenges posed by big data, as far as data storage is concerned, distributed file system (DFS) is being put to use. From the first distributed file system, Google File System (Ghemawat, Gobioff, and Leung 2003), to Hadoop Distributed File System or HDFS (Shvachko et al. 2010), there is a range of solutions available in this category. One of the latest and most popular additions to this category is NoSQL database solution like MongoDB[2] and platforms like Cassandra[3]. This phase of the big data lifecycle contributes to the reliability and availability of data.

4 Processing

The last stage of the big data lifecycle is processing, during which various analytical approaches and methods are implemented on the available data, for basic and advanced analytics. Like traditional data analysis, the objective of big data analysis is extraction of useful information from the available data. Common methods used for this purpose include clustering, classification and data analysis techniques, besides many others. It is important to mention here that traditional processing techniques have to be adapted to the big data scenario by redesigning them for use with parallel computing techniques like the MapReduce programming paradigm (Lee et al. 2012). This phase contributes to the value characteristic of big data. The rest of the chapter will discuss this facet of the big data lifecycle in detail.

[2] https://www.mongodb.com/

[3] http://cassandra.apache.org/

## 4.0 Techniques for Big Data Processing

Big data processing requires a synergistic approach involving mathematical, statistical and optimization techniques, which are implemented using established technologies like data mining, machine learning and signal processing, in addition to several others, for application-specific processing. This makes big data processing essentially inter-disciplinary. An elaborative description of the techniques used for big data processing has been given below.

### *4.1 Mathematical Analysis Techniques*

#### *4.1.1 Mathematical Techniques*

Most big data problems can be mathematically modeled and solved using mathematical analysis techniques like factor analysis and correlation analysis. Factor analysis is mostly used for analysis of relationships between different elements that constitute big data. As a result, it can be used for revealing the most important information. Taking the relationships analysis a step further, correlation analysis can be used for extracting strong and weak dependencies (Ginsberg et al. 2009). Analytics for applications belonging to fields like biology, healthcare, engineering and economics require the use of such techniques.

### *4.1.2 Statistical Methods*

Statistical methods are mathematical techniques that are used for collection, organization and interpretation of data. Therefore, they are commonly used for studying causal relationships and co-relationships. It is also the preferred category of techniques used for deriving numerical descriptions. With that said, the standard techniques cannot be directly implemented for big data. In order to adapt the classical techniques for big data usage, parallelization has been attempted. Research fields related to this area are statistical computing (Klemens 2009), statistical learning (Hastie, Tibshirani, and Friedman 2009) and data-driven statistical techniques (Bennett et al. 2009). The economic and healthcare sectors make extensive use of statistical methods for various applications.

### *4.1.3 Optimization Methods*

Core fields of study like physics, biology and economics involves a lot of quantitative problems. In order to solve these problems, optimization methods are used. Some of these methods that have found wide-ranging use, because of the ease with which they can be parallelized include Genetic Algorithm, Simulated Annealing, Quantum Annealing and Adaptive Simulated Annealing (Sahimi and Hamzehpour 2010).

Nature-inspired optimization techniques like Particle Swarm Optimization and Evolutionary programming have also proven to be useful methods for solving optimization problems. However, these algorithms and techniques are highly storage and computationally intensive. Many research works have attempted scaling of these techniques (Yang, Tang, and Yao 2008, Xiaodong and Xin 2012). An important requirement of big data applications in this regard is real-time optimization, particularly in WSNs (Seenumani, Sun, and Peng 2011).

### *4.2 Data Analytics Techniques*

#### *4.2.1 Data Mining*

Data mining allows extraction of useful information from raw datasets and visualization of the same in a manner that is helpful for making decisions. Commonly used data mining techniques include classification, regression analysis, clustering, machine learning and outlier detection. In order to analyze different variables and how they are dependent on one another, regression analysis may be used. Companies commonly use this technique for analyzing CRM big data and evaluating varied levels of customer satisfaction and its impact on customer retention.

Taking this analysis further, there may also be a need to cluster similar customers together to analyze their buying patterns or classify them on the basis of certain attributes. Clustering and classification are the techniques used for this purpose. Association rule mining may be used for exploring hidden relationships and patterns in big datasets. Lastly, outlier detection is used for fraud detection or risk reduction by identifying patterns or behaviors that are abnormal.

#### *4.2.2 Machine Learning*

A sub-field of artificial intelligence, machine learning allows systems to learn and evolve using empirical data. As a result, intelligent decision-making is fundamental to any system that implements machine learning. However, in the big data context, standard machine learning algorithms need to be scaled up to cope with big data requirements. Deep learning is a recent field that is attracting immense research attention lately (Bengio, Courville, and Vincent 2013). Several Hadoop-based frameworks like Mahout are available for scaling up machine learning algorithms. However, several fields of machine learning like natural language processing and recommender systems, apart from several others, face scalability issues that need to be mitigated for development of generic as well as efficient application-specific solutions.

*4.2.3 Signal Processing*

The introduction of Internet and mobile technologies has made the use of social networking portals and devices like mobiles and sensors, excessively common. As a result, data is being generated at a never-seen-before rate. The massive scale of data available, presence of anomalies, need for real-time analytics and relevance of distributed systems gives rise to several signal processing opportunities in big data (Bo-Wei, Ji, and Rho 2016).

*4.2.4 Neural Networks*

Image analysis and pattern recognition are established applications of Artificial Neural Networks (ANN). It is a well-known fact that as the number of nodes increase; the accuracy of the result gets better. However, the increase in node number elevates the complexity of the neural network, both in terms of memory consumption and computing requirements. In order to combat these challenges, the neural network needs to be scaled using distributed and parallel methods (Mikolov et al. 2011). Parallel training implementation techniques can be used with deep learning for processing big data.

*4.2.5 Visualization Methods*

In order to make the analysis usable for the end user, analytics results need to be visualized in an understandable and clear manner. The high volume and excessive rate of generation of data makes visualization of big data a daunting challenge. Evidently, it is not possible to use traditional visualization methods for this purpose. Most systems, currently available, perform rendering of a reduced dataset to dodge the complications associated with visualization of large datasets (Thompson et al. 2011). However, real-time visualization of big data still presents innumerable challenges.

**5.0 Big Data Computing Models**

Different applications make use of distinctive type, volume and velocity of data. Therefore, the type of computing models applicable depends on the nature of the application. While some applications require batch processing, others need real-time data processing. This section gives taxonomy (see Fig. 3) of big data computing models and discusses the different tools belonging to each of the categories.

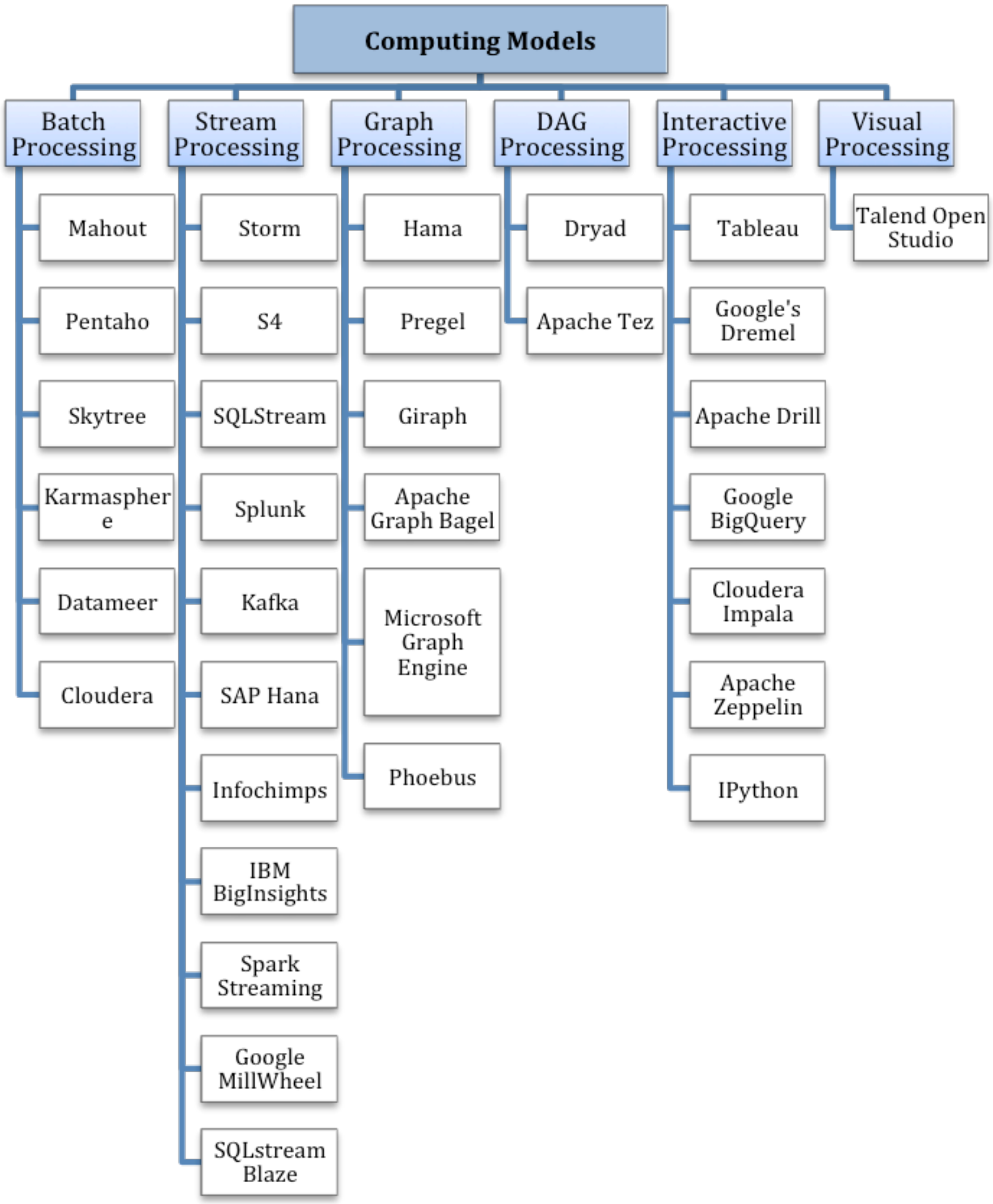

Fig. 3 – Big Data Computing Models

### *5.1 Batch Processing*

Theoretically, batch processing is a processing mode in which a series of jobs are performed on a batch of inputs. MapReduce programming paradigm is the most effective and efficient solution for batch processing of big data. Hadoop, a MapReduce implementation, is identified as the most popular big data processing platform. Therefore, most of the tools described in Table 2 are either Hadoop-based or tools that run on top of Hadoop.

Table 2 – Batch Processing Tools

| Tool | Description | Functionality Supported | Link |
|---|---|---|---|
| Mahout | Used for scalable machine learning in the parallel environment | Capable of performing classification, pattern analysis, clustering analysis, regression analysis and dimension reduction | http://mahout.apache.org/ |
| Pentaho | Hadoop-based software platform for business reports generation | Allows data acquisition, analytics and visualization of business data, in addition to plugin-based communication with NoSQL databases like MongoDB and platforms like Cassandra. | http://www.pentaho.com/product/big-data-analytics |
| SkyTree | A general-purpose server used for machine learning and advanced analytics, which enables optimized machine learning implementation for real-time analytics. | Can be used for similarity search, density estimation, dimension reduction, outlier detection, regression, and classification and clustering | http://www.skytree.net/ |
| Karmasphere | Hadoop-based platform for analysis of business big data | Allows ingestion, iterative analysis and reporting and visualization of business data. | http://www.karmasphere.com/ |
| Datameer Analytic Solution (DAS) | Hadoop-based Platform-as-a-Service (PaaS) solution that allows analysis of business data | Includes data source integration, analytics engine and visualization methods like charts and dashboards. However, data integration is the main focus and it allows data import from structured and unstructured sources. It needs to be deployed in with Cloudera-based Hadoop distribution. | http://www.datameer.com/ |
| Cloudera | Apache Hadoop distribution system | Supports Apache tools like Pig and Hive, along with embedded plugins with Oracle | http://www.cloudera.com/ |

### *5.2 Stream Processing*

In some applications, particularly for real-time analytics, data is available in the form of streams or continuous data flows. These data streams need to be processed, record-by-record, in-memory. Stream processing is considered to be the next-generation computing paradigm for big data. The tools available for this purpose have been described in Table 3.

Table 3 – Stream Processing Tools

| Tool | Description | Functionality Supported | Link |
|---|---|---|---|
| Storm | Open-source, scalable, fault-tolerant, distributed system for | Storm allows programmers to create setup for processing of | http://storm.apache.org/ |
| | real-time computation | streaming data for real-time analytics. It also supports distributed remote procedure call, interactive operating system and online machine learning. | |
| S4 | A scalable, pluggable, distributed, fault-tolerant computing platform | S4 allows development of applications that require effective cluster management and is used by Yahoo for large-scale search query computation. | http://incubator.apache.org/s4/ |
| SQLStream | A platform that supports intelligent, automatic operations, for processing unbounded large-scale data streams | Pattern discovery in unstructured data | http://www.sqlstream.com/blaze/s-server/ |
| Splunk | A platform for analyzing machine-generated data streams | Allows analysis of machine-generated log files, structured as well as unstructured | http://www.splunk.com/ |
| Kafka | Developed for LinkedIn for in-memory management and processing of messaging and stream data | Allows analysis of operational data like service logs and activity data such as user actions for organizational management and usage | http://kafka.apache.org/ |
| SAP Hana | A tool for in-memory processing of data streams | Supports three types of analytics: text analysis and predictive analytics, data | https://hana.sap.com |
| | | warehousing and operational reporting | |
| Infochimps | A cloud suite that provides Infrastructure-as-a-Service (IaaS) provisioning | Supports cloud streams, cloud-based Hadoop and cloud queries for private as well as public clouds, in addition to support for Hive, Pig and Kafka, in addition to several other tools | http://www.infochimps.com/ |
| BigInsights | A stream-based tool by IBM, used in Infosphere platform for big data analytics | Used for real-time analysis of data streams and supports Hive, Pig, Lucene and Oozie, apart from a few others. | http://www-03.ibm.com/software/products/en/ibm-biginsights-for-apache-hadoop |
| Spark Streaming | Stream processing component of Spark | Framework that allows development of fault-tolerant streaming applications. | http://spark.apache.org/streaming/ |
| Google MillWheel (Akidau et al. 2013) | Framework for fault-tolerant stream processing | Supports development of low-latency data processing applications. | - |
| SQLstream Blaze | Stream processing suite | Real-time operational intelligence for stream data analytics of machine data | http://www.sqlstream.com/blaze/ |

### *5.3 Graph Processing*

Owing to the connected nature of data and the significance of exploring relationships and associations for big data analytics, a graph is perhaps the best-suited mathematical model for a majority of the applications. Big data graph processing techniques work in accordance with the Bulk Synchronous Parallel (BSP) computing paradigm (Cheatham et al. 1996). This computing paradigm is commonly used in Cloud Computing. Table 4 gives a list of graph processing architectures and models for graph processing, also elaborating on their features and supported functionalities.

Table 4 – Graph Processing Systems

| **Tool** | **Description** | **Functionality Supported** |
|---|---|---|
| Hama (Seo et al. 2010) | BSP-inspired computing paradigm that runs on top of Hadoop | Used for performing network computations, graph functions, matrix algorithms and scientific computations |
| Pregel (Malewicz et al. 2010) | A graph computation model that approaches problems using BSP programming model | Allows efficient processing of billions of graph vertices that may be connected to each other via trillions of edges |
| Giraph (http://giraph.apache.org/) | Scalable, iterative graph processing system | Adapted from Pregel, Giraph allows edge-oriented input, out-of-core computation and master computation |
| Apache Spark Bagel (Spark 2016) | A Pregel implementation, which is a component of Spark | Supports combiners, aggregators and other basic graph computation |
| Microsoft Graph Engine (Microsoft 2016) | In-memory and distributed graph processing engine | Allows high-throughput offline analytics and low-latency online query processing for very large graphs |
| Phoebus (XSLogic 2016) | Large-scale graph processing | |
| | framework | |

### *5.4 DAG Processing*

In theory, DAG or Directed Acyclic Graph is described as a finite and directed graph. However, this graph lacks any directed cycles. A DAG processing system represents jobs in the system as vertices of the DAG, which execute in parallel. DAG processing is considered a step-up from the MapReduce programming model as it avoids the scheduling overhead prevalent in MapReduce and provides developers with a convenient paradigm for modeling complex applications that require multiple execution steps. Dryad (Isard et al. 2007) is based on dataflow graph processing-based programming model that supports scalable and distributed programming applications. Another application framework that allows execution of a complex DAG created from tasks is Apache Tez[4]. This framework is built on top of YARN.

### *5.5 Interactive Processing*

There needs to be a system that sits between big data applications and users to facilitate smooth communication between the two entities, in order to make big data applications usable. For this purpose, large-scale data processing systems that make use of an interactive mechanism are put to use. Some of the tools that are developed for this purpose have been mentioned in Table 5.

Table 5 – Interactive Processing Tools

| Tool | Description | Functionality Supported | Link |
|---|---|---|---|
| Tableau | Hadoop-based visualization environment that makes use of | Includes three versions: Tableau Desktop (visualization and | http://www.tableau.com/ |

[4] https://tez.apache.org/

| | | | |
|---|---|---|---|
| | Apache Hive to process queries | normalization of data), Tableau Server (business intelligence system for browser-based analytics), and Tableau Public (creating interactive visualizations) | |
| Google Dremel (Melnik et al. 2011) | Scalable interactive analysis system | Performs nested data processing and allows querying of very large tables | - |
| Apache Drill | Interactive analysis system inspired by Dremel, which stores data in HDFS and processes data using MapReduce | Performs nested data processing and supports many data sources and types and different queries. | https://www.mapr.com/products/apache-drill |
| Google BigQuery | An interactive analysis framework that implements Dremel | Uses the Google infrastructure and queries are made on 'append-only' tables, allowing super-fast SQL queries. | https://cloud.google.com/bigquery/ |
| Cloudera Impala | A interactive analysis framework that is inspired by Dremel | Capable of supporting high-concurrency workloads and is the true native interactive solution for Hadoop. | https://www.cloudera.com/products/apache-hadoop/impala.html |
| Apache Zeppelin | Interactive data analytics provisioned in the form of a web-based notebook | Integrates with Spark for seamless data ingestion, discovery, analytics and visualization | https://zeppelin.apache.org/ |
| IPython | Interactive computing architecture | Interactive shell that works as | https://ipytho |
| | | the kernel for Jupyter[5] and supports interactive data visualization. | n.org/ |

### *5.6 Visual Processing*

Visualization is the last and most critical phase of any big data application. This is particularly the case for applications where end-users are involved, simply because information not communicated properly is information not communicated at all. One of the most popular tools for visual big data analytics is Talend Open Studio[6]. It can be integrated with Hadoop and user interfaces can be easily created using its drag-and-drop functionality. Moreover, it also offers RSS feed functionality.

## 6.0 Cloud Computing for Big Data Analytics

John McCarthy gave the world the concept of 'utility computing', during the MIT Centennial talk 1961, when he openly spoke about the future of this industry and how computing will share screen with utilities like electricity and water, and be served as such (Qian et al. 2009). Since then, Cloud Computing has come a long way to become the technology that transforms McCarthy's ideas into reality. However, it was not until 2006 that this technology reached the commercial arena. The introduction of solutions like Amazon's Elastic Compute Cloud (Amazon EC2[7]) and Google App Engine[8] have been historic milestones in the history of Cloud

---

[5] https://jupyter.org/

[6] https://www.talend.com/products/talend-open-studio

[7] https://aws.amazon.com/ec2/

[8] https://appengine.google.com/

Computing.

One of the most initial definitions of Cloud Computing was given by Gartner, which described this technology as a computing style that delivers scalable and elastic IT-enabled capabilities to customers, as a service, with the support of Internet technologies (Gartner n.d.). This definition could not be considered a standard for the sheer simplicity and ambiguity that it entails. National Institute of Standards and Technology (NIST) gave the industry-standard definition for Cloud Computing (Mell and Grance 2011).

According to this definition, Cloud Computing is a technology that allows on-demand, convenient and ubiquitous network access to computing resources that can be configured with minimal requirement of management and interaction with the service provider. The NIST definition also mentioned the five key characteristics, deployment models and delivery models for Cloud Computing. An overview of the NIST definition of Cloud Computing is illustrated in Fig. 4. There are three main components of the Cloud Computing Ecosystem namely, end-user or consumer, distributed server and data center. The cloud provider provisions the IT resources to the end-user with the help of distributed server and data center.

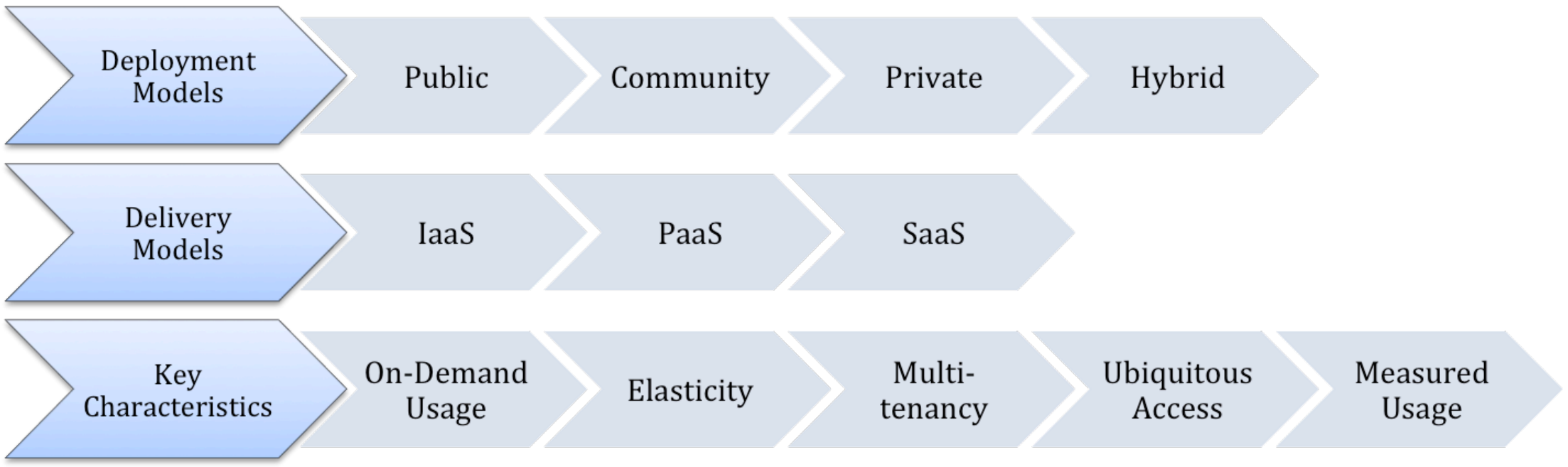

Fig. 4 – Cloud Computing Model Defined By NIST

### *6.1 Key Characteristics*

The five key characteristics of Cloud Computing are as follows -

- On-Demand Usage

  The freedom, to self-provision the cloud-based IT resources, lies in the hands of the end-user.

- Elasticity

  Some users may require dynamic scaling of IT resources, assigned to them. This scaling must be performed on the basis of runtime conditions. Cloud computing's support for this functionality makes it a cost-effective IT provisioning system.

- Multi-tenancy

  Multi-tenancy is a technology that makes use of virtualization to assign and reassign resources to consumers who are isolated from each other.

- Ubiquitous Access

  The end users or consumers of the provided cloud services may have different needs, particularly with respect to their ability to use the self-provisioned IT resources. They may use these resources on any device and using any interface. Therefore, the cloud service must be widely accessible.

- Measured Usage

  In the Cloud Computing paradigm, the consumer only pays for the services he or she uses. To enable this, the usage data related to each consumer needs to be measured. Billing is done on the basis of this measurement. Besides this, these measurements are also used for monitoring purposes and possess a close relationship with on-demand provisioning.

### *6.2 Delivery Models*

It is evident from the discussion that the cloud provider offers services to the consumers.

Obviously, these IT services need to be packaged into combinations to make self-provisioning possible. This pre-packaging of services is called the cloud delivery model. There are three main delivery models used as standards, which have been described below. However, several other delivery models have been created on the basis of applications and functionality that they support. Some examples of these models include Database-as-a-Service, Security-as-a-Service and Testing-as-a-Service, apart from many others.

*6.2.1 Infrastructure-as-a-Service (IaaS)*

In this delivery model, cloud-based tools and interfaces allow access and management of infrastructure-centric IT resources. Therefore, reserves like hardware, network, connectivity and raw IT resources are included in this model. The user is free to configure these resources in the manner that he or she desires. Popular cloud providers of IaaS are Rackspace[9] and Amazon EC[2].

*6.2.2 Platform-as-a-Service (PaaS)*

While infrastructure forms the base layer and requirement for any kind of development or usage, developers may also require pre-deployed and pre-configured IT resources. This gives them a complete environment to directly work upon, which saves time and effort. Some of the most popular products in this category are .NET-based environment provided by Microsoft Azure[10] and a Python-based environment by the Google App Engine.

*6.2.3 Software-as-a-Service (SaaS)*

The shared cloud service can also host software solutions that can directly be used by consumers on need basis. Products like Google Docs[11], which is a shared service and provisions

---

[9] https://www.rackspace.com

[10] https://azure.microsoft.com/

[11] https://www.google.co.in/docs/about/

documentation software and storage to consumers, are examples of SaaS.

*6.2.4 Hybrid Models*

Apart from the three formal models used for delivery of cloud solutions, the user also has the option to use any combinations of these models. In fact, all the three models may also be combined and provisioned to the end-user.

## *6.3 Deployment Models*

The cloud environment has three main characteristics namely, size, ownership and access. On the basis of these three characteristics, the deployment model for the cloud environment is determined.

*6.3.1 Public Cloud*

In such a cloud environment, the access is open to all and a third-party cloud provider owns the cloud. Therefore, creation and maintenance of the cloud environment is the sole responsibility of the cloud provider. The size of the cloud is large owing to the 'accessible to everyone' nature of it.

*6.3.2 Community Cloud*

The Community Cloud is an adaptation of the Public Cloud. The only difference between these two types of deployments is that the community cloud restricts access to the cloud services. Only a small community of cloud consumers can use the services of this cloud. Moreover, such a cloud may involve joint ownership of the community and a third party cloud provider.

*6.3.3 Private Cloud*

When an individual or organization owns a cloud and limits the access of the cloud to the members of the organization only, the deployment model is known as Private Cloud. The size of such a cloud is much smaller and staff is employed for performing administration and management of the cloud. It is also important to mention that in such a deployment model, the cloud provider and consumer, both are the owners of the cloud.

#### *6.3.4 Hybrid Cloud*

A deployment model created using two or more deployment models is called a hybrid cloud.

#### *6.3.5 Other Deployment Models*

Some other cloud deployment models like External Cloud, Virtual Private Cloud or Dedicated Cloud and Inter-cloud also exist.

### ***6.4 What makes Cloud Computing an Ideal Match for Big Data?***

Big data solutions have two fundamental requirements. The size of the data is 'big'. Therefore, in order to store this data, a large and scalable storage space is required. Moreover, the standard analytics algorithms are computing-intensive. Therefore, an infrastructural solution that can support this level of computation is needed. The cloud meets both these requirements well.

Firstly, there are several low-cost storage solutions available with the cloud. Besides this, the user pays for the services he or she uses, which makes the solution all the more cost effective. Secondly, cloud solutions offer commodity hardware, which allows effective and efficient processing of large datasets. It is because of these two reasons that Cloud Computing is considered an ideal infrastructural solution for big data analytics.

### ***6.5 Hadoop on the Cloud***

One of the most popular frameworks used for big data computing is Hadoop. It is an implementation of MapReduce that allows distributed processing of large, heterogeneous datasets. There are many solutions that allow moving of Hadoop to the cloud. Some of the popular solutions are the ones provided by Amazon's Elastic MapReduce[12] and Rackspace. There are several reasons why running Hadoop on the cloud is gaining immense popularity, some of which have been listed below.

- Cost Effectiveness

  Running any software on the cloud reduces the cost of the operation considerably. Therefore, this model provides a cost-effective solution for Hadoop-based applications.

- Scalable Resource Procurement

  One of the key benefits of the cloud paradigm is that it allows scalable provisioning of resources. This feature is best put to use by the Hadoop framework.

- Support for Efficient Batch Processing

  Hadoop is a batch-processing framework. In other words, jobs submitted to Hadoop are collected and sent for execution at a fixed and temporal basis. Therefore, the resources are not loaded with work all the time. The cloud allows on-demand resource provisioning. Therefore, resources can be procured and elastically increased and decreased in capability depending on the phase of Hadoop operation, making the solution cost-effective yet efficient.

- Handles Variable Requirements of Hadoop Ecosystem

  Different jobs performed by Hadoop vary in the quality and quantity of resources they require. For instance, some jobs require I/O bandwidth while others need more memory support. Whenever a physical setup for Hadoop is created, homogenous machines that can support the highest capabilities are used, resulting in wastage of resources. The cloud solution allows formation of a cluster of heterogeneous machines to support the heaviest

[12] https://aws.amazon.com/elasticmapreduce/

tasks yet optimize usage.

- Processing Data Where It Resides

  Most of the organizations keep their data on the cloud. Therefore, processing it on the cloud makes more sense. Migrating the data to a different execution environment will not only be time-consuming, but it will also be inefficient.

The cloud can be used for creation of different types of clusters, with each component of the cluster having a different configuration and characteristics. The available options for running Hadoop in the Cloud environment have been described below.

*6.5.1 Deploying Hadoop in Public Cloud*

Providers like Hortonworks[13], Cloudera[14] and BigInsights[15] offer Hadoop distributions, which can be deployed and run on public clouds provided by Rackspace, Microsoft Azure and Amazon Web Services[16]. Such a configuration is typically referred to as 'Hadoop-as-a-Service'. The issue with such solutions is that they use Infrastructure-as-a-Service (IaaS) provided by the cloud providers. In other words, the IT resources being used are shared between many customers. This gives the user little control over the configuration of the cluster. As a result, there is no concept of rack awareness that the user can configure and access. Besides this, the availability and performance of the cluster are also dependent on the VM (Virtual Machine) that is being used.

The user is required to install and configure Hadoop on all of the available options, other than Amazon EMR. Amazon EMR provides what can be referred to as 'MapReduce-as-a-Service'.

---

[13] http://hortonworks.com/

[14] http://www.cloudera.com/

[15] http://www-03.ibm.com/software/products/en/ibm-biginsights-for-apache-hadoop

[16] https://aws.amazon.com/

The users can directly run MapReduce jobs on the Amazon EMR-powered cluster, making the whole exercise fairly simple and easy for the developer. If the developer does not wish to use HDFS as the default storage solution, then Hadoop cluster can also be used with S3[17]. Although, S3 is not as efficient as HDFS, it provides some unparalleled features like data loss protection, elasticity and bucket versioning. Some applications may require the implementation of these features, which makes S3 an irreplaceable storage solution. Besides this, Hadoop with S3 may be the storage solution of choice for organizations that already have their data stored on S3. A commercially available solution that uses this configuration is Netflix.

*6.5.2 Deploying Hadoop in Private Cloud*

The private cloud allows the user to have better control over the configuration of Hadoop in the cloud. Several cloud providers like IBM's SmartCloud[18] for deployment of InfoSphere BigInsights and IBM's PureData[19] System offer PaaS solutions, which provide a pre-built setup for convenient deployment of Hadoop. The advantages of using Hadoop on a private cloud are as follows:

- Better control and visibility of the cluster
- Better mitigation of data privacy and security concerns

*6.5.3 Key Considerations for Deployment*

There are obvious advantages of running Hadoop on the Cloud. However, it is important to understand that this does not come without problems and potential issues. Some of the things that must be paid heed to before using Hadoop on the cloud are given below.

---

[17] https://aws.amazon.com/s3/getting-started/

[18] http://www-03.ibm.com/software/products/en/ibmsmarnote

[19] https://www-01.ibm.com/software/data/puredata/

- The security provided by the Hadoop cluster is very limited in its capability. Therefore, the security requirements and criticality of data being shared with the Hadoop cluster need to be carefully examined, in advance.
- Typically, Hadoop runs on Linux. However, Hortonworks also provides a Hadoop distribution that works with Windows and is available on Microsoft's Azure Cloud. It is important to identify the operating system requirements and preferences before choosing a Cloud-based Hadoop solution.
- Hadoop can never be viewed as a standalone solution. When it comes to designing big data analytics applications, you will need to look beyond the Hadoop cluster and see if the cloud solution supports visualization tools like Tableau and R, to serve your purpose in totality.
- An important consideration that is usually overlooked is data transmission. Is the data already on the cloud or will it have to be loaded from an internal system? If the application needs transferring of data from one public cloud to another, some transmission fees may apply.
- Using the VM-based Hadoop cluster may suffer from performance issues. These arrangements are good solutions only for development and testing purposes or unless performance is not an issue.

## 7.0 Challenges and Opportunities

Big data computing requires the use of several techniques and technologies. MapReduce and Hadoop are certainly the most popular and useful frameworks for this purpose. Apart from Cloud Computing, it has also been proposed that bio-inspired computing; quantum computing and granular computing are potential technologies for big data computing (Philip Chen and Zhang 2014). However, each of these technologies needs to be adapted for this purpose and is not free from potential challenges. Challenges specific to big data computing using these technologies is beyond the scope of this chapter.

Owing to the elasticity and scalability of cloud solutions, this technology is one of the front-runners for big data computing (Talia 2013). With that said, the feasibility and viability of using a synergistic model is yet to be explored. NESSI presented challenges specific to implementation of existing machine learning techniques for big data computing and development of analytics solutions, mentioning the following requirements as fundamental (NESSI 2012).

- There is a need for development of solid scientific foundation, to facilitate selection of method or technique that needs to be chosen.
- There is a need for development of scalable and efficient algorithms that can be used.
- The developed algorithms cannot be implemented unless appropriate technological platforms have been selected.
- Lastly, the solution's business viability must be explored.

Analytics solutions are expected to be uncomplicated and simple despite the complexity of the problem and the challenges involved. Besides this, distributed systems and parallel computing seems to be an appropriate solution to the big data problem. Therefore, the designed solution has to be inclined towards computational paradigms that work on these foundations. There are several identified challenges related to the use of Cloud Computing for big data analytics (Hashem et al. 2015, Assunção et al. 2015). Big data in the cloud suffers from several trials, both technical as well as non-technical. Technical challenges associated with cloud-based big data analytics can further be divided into three categories namely, big data management, application modeling and visualization.

Challenges associated with characteristics, storage and processing of big data are included in big data management. Management of big data is a challenging task considering the fact that data is continuously increasing in volume. Moreover, aggregation and integration of unstructured data, collected from diverse sources is also under research consideration. There are two aspects of data acquisition and integration. Firstly, data needs to be collected from different sources. Besides this, the collected data may be structured, unstructured or semi-structured, in type. Integration of a variety of data types into an aggregation that can be further used for analytics is an even bigger problem.

In order to make predictions, recommendations or implement any application-specific functionality, the application needs to be modeled. Moreover, this model needs to evolve with the changing needs of the system. The systems of today are generating both static as well as stream data. Therefore, any solution designed for such a system must support integration of different programming models to form a generic analytics engine. It is not possible to create practically viable solutions unless these models make use of resources optimally and exhibit high energy-efficiency, which adds to the secondary requirements of the system. The expected future work with respect to the analytics engine can be summarized as,

- Development of energy-efficient and cost-effective solutions
- Standardization of solutions

The findings of the analytics engine need to be presented to the user in a manner that the results can be of use. This makes visualization an invincible part of any analytics solution. Better and more cost-effective hardware support for large-scale visualization is being investigated (Assunção et al. 2015). In addition to this, there is a pressing need for solutions that can visualize big data efficiently. One of the most recent and popular research topics in visualization is real-time visualization. Since systems are generating data and analytics results in real-time, presenting the same to the user in real-time makes more sense. Visualization in the cloud context suffers from an inherent issue. Network, for such applications, has long been and continues to remain a bottleneck.

Technology needs to be practically viable to support high business value solutions. With that said, cloud-based big data technologies suffers from several non-technical challenges that need to be worked out before widespread adoption can be expected. The lack of generic and standard solutions affects the business value of this technology heavily. Besides this, even though the use of Cloud Computing has brought down the cost of big data technologies considerably, more cost-effective models that make optimal use of the elasticity property of the cloud need to be devised for the end-users.

Businesses find it difficult to migrate to big data technologies due to lack of skilled staff and debugging-testing solutions available for these technologies. One of the biggest concerns while using big data analytics and Cloud Computing in an integrated model is security. In view of this, additional challenges like security and privacy risks also exist and need to be mitigated before commercially viable, practical cloud-based big data analytics solutions can be developed and used.

## 7.0 Future Research Directions

In view of the challenges and issues identified in the previous section, there is a need for a Cloud-based framework to facilitate advanced analytics. The analytical workflow is composed of several steps, which include data acquisition, storage, processing, analytics and visualization.

Besides this, each of these steps is composed of many sub-steps. For instance, data acquisition is composed of sub-steps like data collection, pre-processing and transformation.

With specific focus on processing and analytics, existing solutions in the field are not generic. There is tight coupling between field-specific analytical solutions and data model used for the same. In addition to this, data model diversity also exists as a fundamental issue for generic framework development. Lastly, the data characteristics used to classify data as big data varies substantially with time and changes from one application to another. Research needs to be directed towards development of a generic analytical framework and cloud-based big data stack that addresses the complexities of issues mentioned above.

Big data technology applies and appeals to every walk of human life. There is no technology-enabled system that cannot make use of the big data-powered solutions for enhanced decision making and industry-specific applications. However, in order to make this technology commercially viable, research groups need to identify potential 'big' datasets and possible analytical applications for the field concerned. With that said, the feasibility and commercial viability of such analytical applications need to aligned with business and customer requirements.

## References

Abu-Mostafa, Yaser S., Malik Magdon-Ismail, and Hsuan-Tien Lin. 2012. *Learning From Data*. United States: AMLBook.com.

Akidau, Tyler, Sam Whittle, Alex Balikov, Kaya Bekiroğlu, Slava Chernyak, Josh Haberman, Reuven Lax, Sam McVeety, Daniel Mills, and Paul Nordstrom. 2013. "MillWheel." *Proceedings of the VLDB Endowment* 6 (11):1033-1044. doi: 10.14778/2536222.2536229.

Assunção, Marcos D., Rodrigo N. Calheiros, Silvia Bianchi, Marco A. S. Netto, and Rajkumar Buyya. 2015. "Big Data computing and clouds: Trends and future directions." *Journal of Parallel and Distributed Computing* 79-80:3-15. doi: 10.1016/j.jpdc.2014.08.003.

Bengio, Y., A. Courville, and P. Vincent. 2013. "Representation learning: a review and new perspectives." *IEEE Trans Pattern Anal Mach Intell* 35 (8):1798-828. doi:

10.1109/TPAMI.2013.50.
Bennett, Janine, Ray Grout, Philippe Pebay, Diana Roe, and David Thompson. 2009. "Numerically stable, single-pass, parallel statistics algorithms."1-8. doi: 10.1109/clustr.2009.5289161.
Bo-Wei, Chen, Wen Ji, and Seungmin Rho. 2016. "Divide-and-conquer signal processing, feature extraction, and machine learning for big data." *Neurocomputing* 174:383. doi: 10.1016/j.neucom.2015.08.052.
Burnap, Peter, Omer Rana, Matthew Williams, William Housley, Adam Edwards, Jeffrey Morgan, Luke Sloan, and Javier Conejero. 2014. "COSMOS: Towards an integrated and scalable service for analysing social media on demand." *International Journal of Parallel, Emergent and Distributed Systems* 30 (2):80-100. doi: 10.1080/17445760.2014.902057.
Buyya, Rajkumar. 2016. "Big Data Analytics = Machine Learning + Cloud Computing." In *Big Data*, 7-9. Massachusetts, USA: Morgan Kaufmann Publisher.
Chandio, Aftab Ahmed, Nikos Tziritas, and Cheng-Zhong Xu. 2015. "Big-data processing techniques and their challenges in transport domain." *ZTE Communications*.
Cheatham, Thomas, Amr Fahmy, Dan Stefanescu, and Leslie Valiant. 1996. "Bulk Synchronous Parallel Computing — A Paradigm for Transportable Software."61-76. doi: 10.1007/978-1-4615-4123-3_4.
Gartner. 2016. "Gartner IT Glossary." Gartner Inc.
Ghemawat, Sanjay, Howard Gobioff, and Shun-Tak Leung. 2003. "The Google file system." *ACM SIGOPS Operating Systems Review* 37 (5):29. doi: 10.1145/1165389.945450.
Ginsberg, J., M. H. Mohebbi, R. S. Patel, L. Brammer, M. S. Smolinski, and L. Brilliant. 2009. "Detecting influenza epidemics using search engine query data." *Nature* 457 (7232):1012-4. doi: 10.1038/nature07634.
Hammond, Klavdiya, and Aparna S. Varde. 2013. "Cloud Based Predictive Analytics: Text Classification, Recommender Systems and Decision Support."607-612. doi: 10.1109/icdmw.2013.95.
Hashem, Ibrahim Abaker Targio, Ibrar Yaqoob, Nor Badrul Anuar, Salimah Mokhtar, Abdullah Gani, and Samee Ullah Khan. 2015. "The rise of “big data” on cloud computing: Review and open research issues." *Information Systems* 47:98-115. doi: 10.1016/j.is.2014.07.006.
Hastie, Trevor, Robert Tibshirani, and Jerome Friedman. 2009. "The Elements Of Statistical Learning " *Springer Series In Statistics*. doi: 10.1007/978-0-387-84858-7.
Isard, Michael, Mihai Budiu, Yuan Yu, Andrew Birrell, and Dennis Fetterly. 2007. "Dryad." *ACM SIGOPS Operating Systems Review* 41 (3):59. doi: 10.1145/1272998.1273005.
Khan, Zaheer, Ashiq Anjum, and Saad Liaquat Kiani. 2013. "Cloud Based Big Data Analytics for Smart Future Cities."381-386. doi: 10.1109/ucc.2013.77.
Klemens, Ben. 2009. *Modeling With Data*. Princeton, N.J.: Princeton University Press.
Lee, Kyong-Ha, Yoon-Joon Lee, Hyunsik Choi, Yon Dohn Chung, and Bongki Moon. 2012. "Parallel data processing with MapReduce." *ACM SIGMOD Record* 40 (4):11. doi: 10.1145/2094114.2094118.
Lu, Sifei, Reuben Mingguang Li, William Chandra Tjhi, Kee Khoon Lee, Long Wang, Xiaorong Li, and Di Ma. 2011. "A Framework for Cloud-Based Large-Scale Data Analytics and Visualization: Case Study on Multiscale Climate Data."618-622. doi: 10.1109/CloudCom.2011.95.

Malewicz, Grzegorz, Matthew H. Austern, Aart J. C. Bik, James C. Dehnert, Ilan Horn, Naty Leiser, and Grzegorz Czajkowski. 2010. "Pregel."135. doi: 10.1145/1807167.1807184.

Mell, P. M., and T. Grance. 2011. doi: 10.6028/nist.sp.800-145.

Melnik, Sergey, Andrey Gubarev, Jing Jing Long, Geoffrey Romer, Shiva Shivakumar, Matt Tolton, and Theo Vassilakis. 2011. "Dremel." *Communications of the ACM* 54 (6):114. doi: 10.1145/1953122.1953148.

Microsoft. 2016. "Graph Engine." accessed 2016.

Mikolov, Tomas, Anoop Deoras, Daniel Povey, Lukas Burget, and Jan Cernocky. 2011. "Strategies for training large scale neural network language models."196-201. doi: 10.1109/asru.2011.6163930.

NESSI. 2012. "Big Data: A New World Of Opportunities." accessed 2016.

Philip Chen, C. L., and Chun-Yang Zhang. 2014. "Data-intensive applications, challenges, techniques and technologies: A survey on Big Data." *Information Sciences* 275:314-347. doi: 10.1016/j.ins.2014.01.015.

Qian, Ling, Zhiguo Luo, Yujian Du, and Leitao Guo. 2009. "Cloud Computing: An Overview." 5931:626-631. doi: 10.1007/978-3-642-10665-1_63.

Raghupathi, W., and V. Raghupathi. 2014. "Big data analytics in healthcare: promise and potential." *Health Inf Sci Syst* 2:3. doi: 10.1186/2047-2501-2-3.

Ratner, Bruce. 2003. *Statistical Modeling And Analysis For Database Marketing*. Boca Raton, Fla.: Chapman & Hall/CRC.

Sahimi, Muhammad, and Hossein Hamzehpour. 2010. "Efficient Computational Strategies for Solving Global Optimization Problems." *Computing in Science & Engineering* 12 (4):74-83. doi: 10.1109/mcse.2010.85.

Salminen, A. 2012. "Introduction to NOSQL." NoSQL Seminar 2012.

Samuel, A. L. 1959. "Some Studies in Machine Learning Using the Game of Checkers." *IBM Journal of Research and Development* 3 (3):210-229. doi: 10.1147/rd.33.0210.

Seenumani, Gayathri, Jing Sun, and Huei Peng. 2011. "Real-Time Power Management of Integrated Power Systems in All Electric Ships Leveraging Multi Time Scale Property." *IEEE Transactions on Control Systems Technology*. doi: 10.1109/tcst.2011.2107909.

Seo, Sangwon, Edward J. Yoon, Jaehong Kim, Seongwook Jin, Jin-Soo Kim, and Seungryoul Maeng. 2010. "HAMA: An Efficient Matrix Computation with the MapReduce Framework."721-726. doi: 10.1109/CloudCom.2010.17.

Shvachko, Konstantin, Hairong Kuang, Sanjay Radia, and Robert Chansler. 2010. "The Hadoop Distributed File System."1-10. doi: 10.1109/msst.2010.5496972.

Spark. 2016. "Spark Documentation." Spark.Apache.Org.

Sun, N., J. G. Morris, J. Xu, X. Zhu, and M. Xie. 2014. "iCARE: A framework for big data-based banking customer analytics." *IBM Journal of Research and Development* 58 (5/6):4:1-4:9. doi: 10.1147/jrd.2014.2337118.

Talia, Domenico. 2013. "Clouds for Scalable Big Data Analytics." *Computer* 46 (5):98-101. doi: 10.1109/mc.2013.162.

Thaduri, Adithya, Diego Galar, and Uday Kumar. 2015. "Railway Assets: A Potential Domain for Big Data Analytics." *Procedia Computer Science* 53:457-467. doi: 10.1016/j.procs.2015.07.323.

Thompson, David, Joshua A. Levine, Janine C. Bennett, Peer-Timo Bremer, Attila Gyulassy, Valerio Pascucci, and Philippe P. Pebay. 2011. "Analysis of large-scale scalar data using hixels."23-30. doi: 10.1109/ldav.2011.6092313.

Xiaodong, Li, and Yao Xin. 2012. "Cooperatively Coevolving Particle Swarms for Large Scale Optimization." *IEEE Transactions on Evolutionary Computation* 16 (2):210-224. doi: 10.1109/tevc.2011.2112662.

Xindong, Wu, Zhu Xingquan, Wu Gong-Qing, and Ding Wei. 2014. "Data mining with big data." *IEEE Transactions on Knowledge and Data Engineering* 26 (1):97-107. doi: 10.1109/tkde.2013.109.

XSLogic. 2016. "XSLogic/Phoebus." Github.

Yang, Zhenyu, Ke Tang, and Xin Yao. 2008. "Large scale evolutionary optimization using cooperative coevolution." *Information Sciences* 178 (15):2985-2999. doi: 10.1016/j.ins.2008.02.017.